\documentclass[10pt,a4paper]{IEEEtran} 

\usepackage{url}
\usepackage[utf8]{inputenc}

\usepackage{verbatim}
\usepackage{tabularx}
\usepackage{multirow}
\usepackage{microtype}
\usepackage[super]{nth}

\usepackage{tikz}

\newcommand{\pie}[1]{%
\begin{tikzpicture}
 \draw (0,0) circle (1ex);\fill (1ex,0) arc (0:#1:1ex) -- (0,0) -- cycle;
\end{tikzpicture}%
}

\begin{document}

\title{Scaling Agile Development in Mechatronic Organizations -- A Comparative Case Study}

\author{\IEEEauthorblockN{Ulrik Eklund}
\IEEEauthorblockA{Department of Computer Science and Media Technology\\
Malmö University\\
Malmö, Sweden\\
Email: ulrik.eklund@mah.se}\\

\and
\IEEEauthorblockN{Christian Berger}
\IEEEauthorblockA{Department of Computer Science and Engineering\\
University of Gothenburg\\
Gothenburg, Sweden\\
Email: christian.berger@gu.se}
}

\maketitle



\begin{abstract}
Agile software development principles enable companies to successfully and quickly deliver software by meeting their customers' expectations while focusing on high quality. Many companies working with pure software systems have adopted these principles, but implementing them 
in companies dealing with non-pure software products is challenging.

We identified a set of goals and practices 
to support 
large-scale agile development in companies that develop software-intense mechatronic systems.

We used an inductive approach based on empirical data collected during a
longitudinal study with six companies in the Nordic region. The data collection
took place over two years through focus group workshops, individual on-site
interviews, and complementary surveys.

The primary benefit of large-scale agile development is improved quality, enabled by practices that support regular or continuous integration between teams delivering software, hardware, and mechanics. 
In this regard, the most beneficial integration cycle for deliveries is every four weeks; while continuous integration on a daily basis would favor software teams, other disciplines does not seem to  benefit from faster integration cycles.

We identified 108 goals and development practices 
supporting agile principles among
the companies, most of them concerned with integration; therefrom, 26 agile practices are unique to the mechatronics domain to support adopting agile beyond pure software development teams. 
16 of these practices are considered as
key enablers, confirmed by our control cases.
\end{abstract}

\begin{IEEEkeywords}
software engineering, 
agile software development, 
mechatronics, 
embedded software, 
system integration, 
testing
\end{IEEEkeywords}

\section{Introduction} 
\label{sec:intro}

Agile software development aims at developing products that better match a
customer's expectations compared to waterfall or stage-gate methods. Typical
characteristics of agile methods are short and fixed periods consisting of
development, integration, and testing, conducted in small teams that communicate actively, both within the software team and with the customer. This flexibility allows a team to continuously reprioritize a product's features based on stakeholder feedback. 

Pure software-driven companies are the typical habitat for adopting agile
with prominent examples being Google, Amazon, or Spotify. The mechatronics
domain, though, where cars are a prime example, are more challenging as
the final product combines software, hardware, and mechanics, with the
involved artifacts being of different natures and contributed from different
disciplines.

We can see two opposing trends affecting R\&D in the mechatronics domain:
Manufacturing and hardware development have long lead-times compared to pure
software products, ranging typically 1-4 years. During the product development process, focus is given to predictability, i.e.~meeting the start-of-production (SOP) with the required mechanical quality, which in practice is achieved by waterfall/stage-gate processes dictating delivery and integration cycles.
In contrast, software development is characterized by increasing speed and being more nimble while keeping quality. This typically enables lead-times of weeks or months, and many agile methods are a response to this. However, there are no established solutions to easily overcome the intersection between the aforementioned trends, but the necessity to resolve them in the mechatronics domain motivates in-depth studies to better serve the changing market's needs and to support industrial decision makers.

In this study, we compared experiences and practices from six internationally
leading companies developing and manufacturing mechatronic systems. The software teams in these companies are already following a number of common agile practices, such as small team sizes, regular stand-up meetings,
cross-functional teams, reprioritization, shared backlog, and sprint lengths of up to four weeks. All involved companies are at the threshold to scale agile principles
beyond their individual software teams to reach out to hardware and mechanics.

\subsection{Problem Statement}
\label{sec:problem}
The two trends above typically result in a situation where individual teams are be able to reprioritise and implement software features in a 2-4 weeks cycle, i.e. are agile, while the overall R\&D process is typically still governed by an overarching stage-gate or V-model~\cite{eklund_applying_2012}. Thus, software deliveries were typically planned in time towards pre-scheduled integration points that are determined by mechanics and manufacturing development. As a result the  benefits typically associated with agile development like short lead-times in launching new or updated products were not perceived by developers.

\subsection{Research Objectives}
The aim of the study is to unveil a list of agile practices that are 
enablers to scale agile beyond software development teams in mechatronic organizations. These practices 
scaling agile principles to also include
hardware and mechanics, neighboring groups, and R\&D departments.
Practitioners from mechatronics organizations who are transforming and adjusting their
internal development processes to  accommodate the trends in Section~\ref{sec:problem}
by following large-scale agile frameworks
such as LESS and SAFe \cite{larman_large-scale_2016,larman_large-scale_2014,leffingwell_scaled_????,brenner_scaled_2015}
benefit from this list to identify practices 
supporting a large-scale agile transformation.

\subsection{Context and Limitations}

This study compared organizations with the following characteristics:
\begin{itemize}
\item Large mechatronics organizations,
\item Dealing with a large and diverse product portfolio with regular product upgrades, and
\item Where timely manufacturing plays a large role, while
\item There are strong demands on high quality and safety.
\end{itemize}

\subsection{Contributions}

During our study, we could confirm already known facts about scaled agile development,
such as the challenge of coordinating multiple teams, difficulties with managing requirements,
and hanging on to internal silos \cite{dikert_challenges_2016}.
However, we also identified a number of additional challenges and benefits that are new
and unique for software-intense mechatronic systems. 
The final result of the study is a set of 26 practices for agile development, which
are particular to the mechatronics domain, from which 16 are considered as enablers
in addition to well-known practices for large-scale software development to intensify
the adoption of agile beyond pure software development teams.

\subsection{Structure of the Article}

The rest of the article is structured as follows: Section~\ref{sec:method} describes the overall design of our comparative case study and the embodied methods followed by a presentation of the results in Section~\ref{sec:results}. We discuss our findings with respect to related work in Section~\ref{sec:related} before we conclude in
Section~\ref{sec:conclusion}.


\section{Comparative Case Study Design} 
\label{sec:method}

Following is the description of the design of our comparative case study, conducted over two years. 
Our research project was conducted
as part of Software Center\footnote{\url{http://www.software-center.se}},
an interdisciplinary industrial/academic collaboration environment
opened in 2012 with five Swedish universities 
(Chalmers Institute of Technology,
University of Gothenburg, 
Malmö University, 
University of Linköping, and
Mälardalen University), and ten companies 
(Ericsson AB, 
Saab AB, 
Volvo Group,
Volvo Cars, 
Axis Communication AB, 
Tetra Pak, 
Grundfos, 
Jeppesen, 
VeriSure, and 
Siemens).
Software Center's mission is to improve the way software-intense companies develop, deploy, and maintain strongly software-driven products.

\subsection{Research Questions}
\label{sec:rq}
A growing number of companies dealing with software-intense products 
apply agile principles to track a project's progress and the artifacts' quality while reacting agilely on changing customer's needs. However, companies dealing with
mechatronic products where software is driving the innovation for hardware, face
challenges when
implementing agile principles especially beyond the software development teams
as different disciplines with dedicated and lengthily nurtured
engineering principles meet. In our comparative case study, we aim to systematically
investigate the following questions:

\begin{description}
\item[\textit{RQ-1:}] What are expected benefits and
challenges when scaling agile principles beyond software development teams? 
\item[\textit{RQ-2:}] What key enabling practices are considered to scale agile in mechatronics companies outreaching pure software teams?
\end{description}

\subsection{Case and Subjects Selection}

We collaborated with six companies that develop software-driven mechatronic products
during our study. The companies employ between 2,100 and 93,000 employees with between 100 and 1,000 software developers, and up to 3,000 developers from all disciplines.
All companies serve their customers globally with software-intense mechatronic products in yearly volumes between 0.4 to over 16 million units that many of us may even find in our households.

\begin{figure*}[tbp]
\centering
\includegraphics[width=.9\textwidth]{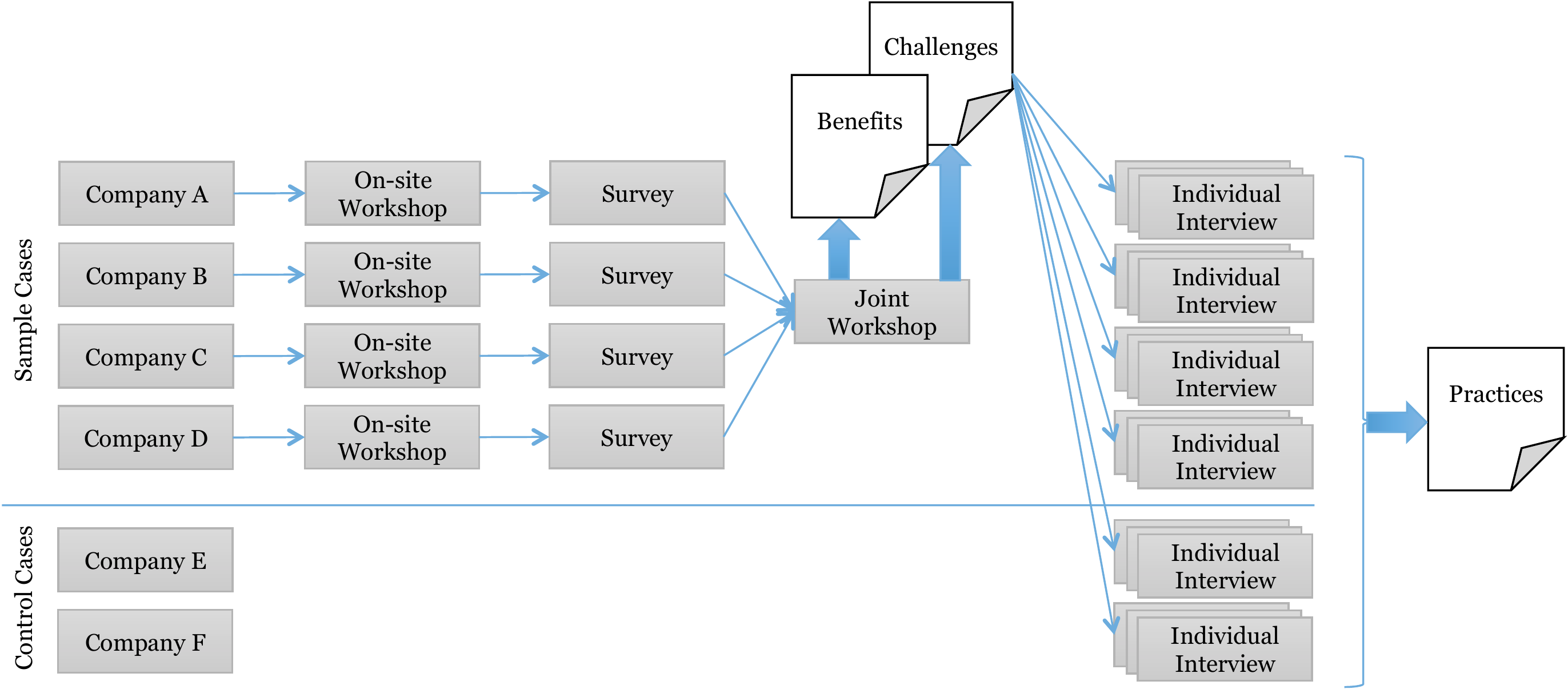}
\caption{Overview of the data collection and analysis procedure: Out of the six participating companies during the project, we selected four companies to extract expected benefits and foreseeable challenges based on on-site workshops, separate surveys per company, and joint workshops involving all four companies. From the identified challenges, we designed a questionnaire that we used to conduct individual interviews with three senior engineers respectively, who specialized in software, hardware, and mechanics per company. We used two companies to confirm findings for agile practices particular to mechatronics.}
\label{fig:dataCollection}
\end{figure*}

The project was defined according to the collaboration model of Software
Center\footnote{\url{http://www.software-center.se/about-software-center}},
with the two researchers defining the initial project with input from four companies,
labeled A-D, sharing a mutual interest in the research topic. Two companies, E and F,
joined the project half-way and acted as control cases, as shown in
Fig.~\ref{fig:dataCollection}.

\subsection{Data Collection Procedure}

We planned the data collection in a multi-stage process as shown in
Fig.~\ref{fig:dataCollection} (a) to get
data from individuals involved in R\&[]D and affected by the software/hardware interplay, (b) to unlock unthought reflections from both, on-site workshops at the involved companies and joint meetings among the involved project partners, and (c) to complement
our data set using surveys with data from further relevant employees.

\subsubsection{Individual On-site Workshops}

In the project's initial phase, we visited four partners individually and organized 3-hours workshops to unveil, collect, and discuss (a) expected benefits and (b) foreseeable challenges when scaling agile beyond software development teams. The number of attendees varied between 5 and 15 people and covered profiles from hardware development, software development, and testing. The participants contributed to the two topics with individual
reflections written on sticky notes. All notes were collected, presented, and jointly discussed with the audience to identify clusters. During the workshops, one researcher acted as moderator while the other researcher took notes and transcribed the sticky notes for further data processing. 

\subsubsection{Complementary Online Surveys per Company}

After the individual on-site workshops, the identified clusters were used
to derive and design a questionnaire. The questionnaire was divided into the
following major sections: Capturing the current situation on adopting agile
principles at the company, the way these practices are perceived at the
respective companies, expected benefits, and expected challenges when scaling
agile\footnote{The questionnaire is available here: \url{https://goo.gl/A58qDf}}.
The goals for the questionnaire were to complement the captured input during
the on-site workshops and to extend the topics that we already identified.
We evaluated the design of the questionnaire with our respective company
points-of-contact before we created separate online surveys for the participating
companies. The online surveys were sent via the respective company
point-of-contacts to further employees.

\subsubsection{Joint Workshop with All Companies}

After conducting the online surveys, we extracted and summarized the feedback
and prepared a joint workshop for all participating companies. We invited an external expert on agile for acting as moderator during these joint workshops.
The external expert's role was defined to stimulate discussions around topics that we might have missed from the individual on-site interviews or in our
online surveys. 
Our role as researchers was to present the results
from the individual workshops and the on-line surveys as well as to take notes
during the discussions.

\subsubsection{Three Individual Interviews with Software, Hardware, and Mechanics Senior Engineers}

The last step in our data collection procedure consisted of individual one-hour interviews per company with one senior engineer each from software, hardware, and mechanics department respectively, who are concerned with
integration and testing.
The goals for these interviews were (a) to identify practices that were perceived supportive for the integration and testing of the respective
three development artifacts, and (b) to unveil aspects that are perceived to slow down the integration and testing. We used the following questions for our
interviews:

\textit{Descriptive Information:}
\begin{enumerate}
\item What is your role?
\item How long have you been working in this role?
\end{enumerate}

\textit{``Let’s assume that you have successfully integrated everything yesterday (software, hardware, and mechanics),\dots''}
\begin{enumerate}
\item How long did that take?
\item How does the integration contribute to your team today?
\item What worked well that you would like to preserve for the next time?
\end{enumerate}

\textit{``Let’s assume that we could integrate again (a) tomorrow, (b) next week, (c) next month, (d) in three months, or (e) in half a year,\dots''}
\begin{enumerate}
\item What would be the value for you and your team?
\item What amount of useful feedback would you be able to get?
\item What would be the cost of doing it again (a) tomorrow \dots (e) in half a year?
\item What resources would be needed to do it again (a) tomorrow \dots (e) in half a year?
\item What would you need to change in order for this to happen?
\item Which other stakeholders at your company could help to achieve a faster integration?
\item Is it preferable for your company to gradually reduce the time between integrations or would you rather make more radical changes?
\end{enumerate}

We concluded the interviews with the following question: \textit{What would you
like to be different compared to the situation of today?}

\subsection{Analysis Procedure}
The data collected from the four different activities covered in total more than 100 pages. The data from the notes was broken down into individual sentences that were transferred to a spreadsheet tool to allow for sorting the
individual statements by data collection phase or company for example, and to cluster the data into topics. We mapped the individual statements through a pre-set coding scheme to two main areas ``process'' and ``product'' and allocated the following development phases to the individual statements: ``requirements engineering'', ``implementation'', ``integration'', ``testing'', and ``deployment''.

We sorted the data by the topics mentioned most often and extracted best or desired practices 
to better facilitate large-scale agile development
As this list contained both practices that are important for 
an agile development process in general, and practices that are only relevant for the mechatronics domain , we annotated those that are specific for the software-intense mechatronics development. 
Furthermore, we indicated per practice what statement from the individual company
interviewees support the extracted statements; the control cases (cf.~Fig.~\ref{fig:dataCollection})
were used to confirm those considered as key-enablers when present in both groups.
Finally, we mapped these identified topics to Stojanov et al.'s agile maturity
model \cite{stojanov_maturity_2015} to allocate the unveiled topics to agile
collaboration phases reflecting an organization's agile maturity level.

\subsection{Validity Procedure}
\label{sec:ValidityProcedure}
We carefully designed the study to reduce bias in data collection and data analysis (cf.~our discussion about Threats to Validity in Sec.~\ref{sec:ThreatsToValidity})
by using method and data triangulation.

\subsubsection{Method Triangulation}
The selection of data collection instruments was part of the validity strategy to minimize bias, and as a result we
used different methods as part of our collection procedure: (a) on-site workshops, (b) questionnaires, (c) individual interviews, (d) joint cross-company feedback and discussion meetings, and (e) joint feedback meetings with the points-of-contact from the respective companies.
During the on-site workshops, we used sticky notes to let participants write their reflections and comments so that we could use their original statements. To validate the logical structure, design, and understandability of the questionnaire, we involved all points-of-contact from the
participating companies to get feedback.
As described before, we invited an external
expert on agile with more than ten years of experience on transforming large-scale organizations to adopt agile principles to act as moderator for the joint workshops with the companies where the researchers acted as observers only.
After every step, our summaries and intermediate findings have been presented and discussed with all points-of-contact.

\subsubsection{Data Triangulation}
Another validity procedure was data triangulation;  we included different data sources during our data collection process to complement
the gathered samples and to increase the chance of getting a more complete picture
of the situation at the various companies. This  identified prevailing practices common
to mechatronics organizations when scaling agile principles. First, we
collaborated with six different mechatronics companies from various product domains having experience with agile software development, but all at the threshold for scaling agile beyond
software development teams. 

Second, to extend and complement the data points from our individual on-site workshops, we reached out to further employees at the respective
companies using separate online surveys. Third, we conducted separate one hour individual interviews with senior engineers from software, hardware, and mechanics to unveil further data especially focusing on 
coordination and integration
between the different disciplines. Finally, we used two of the six companies as control cases as they were not involved in the initial joint workshops.


\section{Results} 
\label{sec:results}

From our data collection procedure, we extracted in total 409 individual statements
from the first three activities and 108 practices from the final individual interviews with the senior engineers. 
From the 409 statements, we could see 216 that were related to expected benefits and 193, which were related to foreseeable challenges; further
processing these items resulted in 35 statements that were clearly product-related and 279 were process-related. The rest  of the statements were personal reflections by the participants on e.g., their background, previous experiences, and the workshop format.

While the participants mentioned nearly three times as many benefits as challenges when
scaling agile for 
their products (26 vs.~9 statements), the results for the organizations' processes are more balanced (129 statements regarding benefits vs.~150 statements regarding challenges).
Further investigating the statements regarding the development phases, we extracted the
distribution as depicted in Tab.~\ref{tab:distributionBenefitsChallenges}

\begin{table}[t]
\centering
\caption{Distribution of statements regarding expected benefits and foreseeable challenges when scaling agile principles beyond the software development related to different development phases: Most challenges are expected during the integration and testing phase.}
\label{tab:distributionBenefitsChallenges}
\begin{tabular}{lcc}
\hline
\textbf{Development Phase} & \textbf{Expect.~Benefits} & \textbf{Fores.~Challenges}\\
\hline
All Phases & 38 & 88 \\
Requirements Engineering & 20 & 9 \\
Implementation & 11 & 6 \\
Integration & 27 & 33 \\
Testing & 17 & 12 \\
Deployment & 16 & 2 \\
\hline
Sum & 129 & 150 \\
\hline
\end{tabular}
\end{table}

From Tab.~\ref{tab:distributionBenefitsChallenges}, no clearly dominating benefit for the
product could be seen in our data. However, it is apparent that most issues are expected from scaling agile in the integration and testing phases, when deliverables from the
different disciplines will be combined for the first time.

\subsection{RQ-1: Benefits and Challenges when Scaling agile}

\begin{figure*}[ht]
\centering
\includegraphics[width=0.74\linewidth]{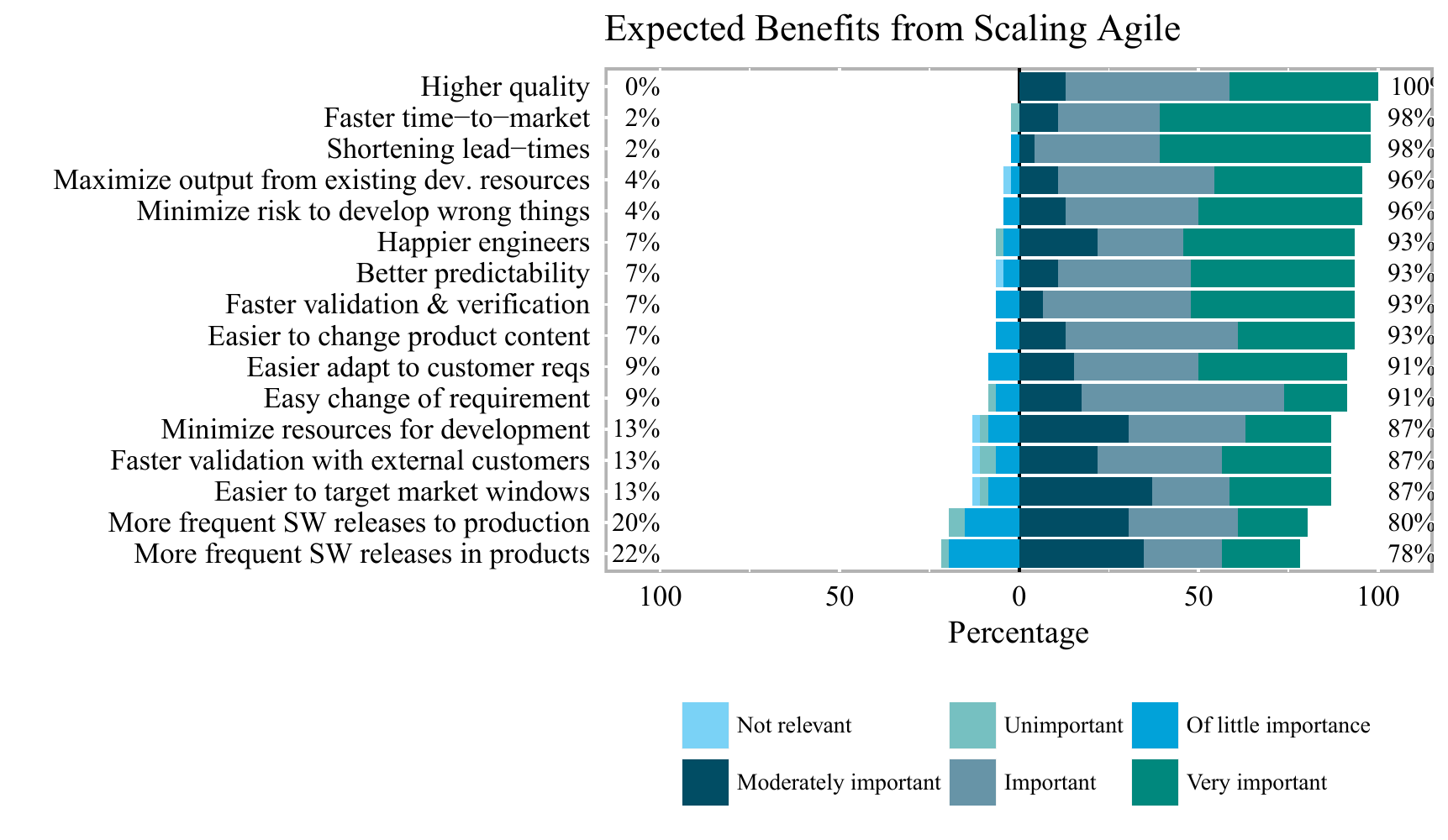}
\caption{Expected benefits when scaling agile beyond software development teams; participants expressed their feedback using an ordered response scale.}
\label{fig:benefits}
\end{figure*}
\begin{figure*}[t]
\centering
\includegraphics[width=0.74\linewidth]{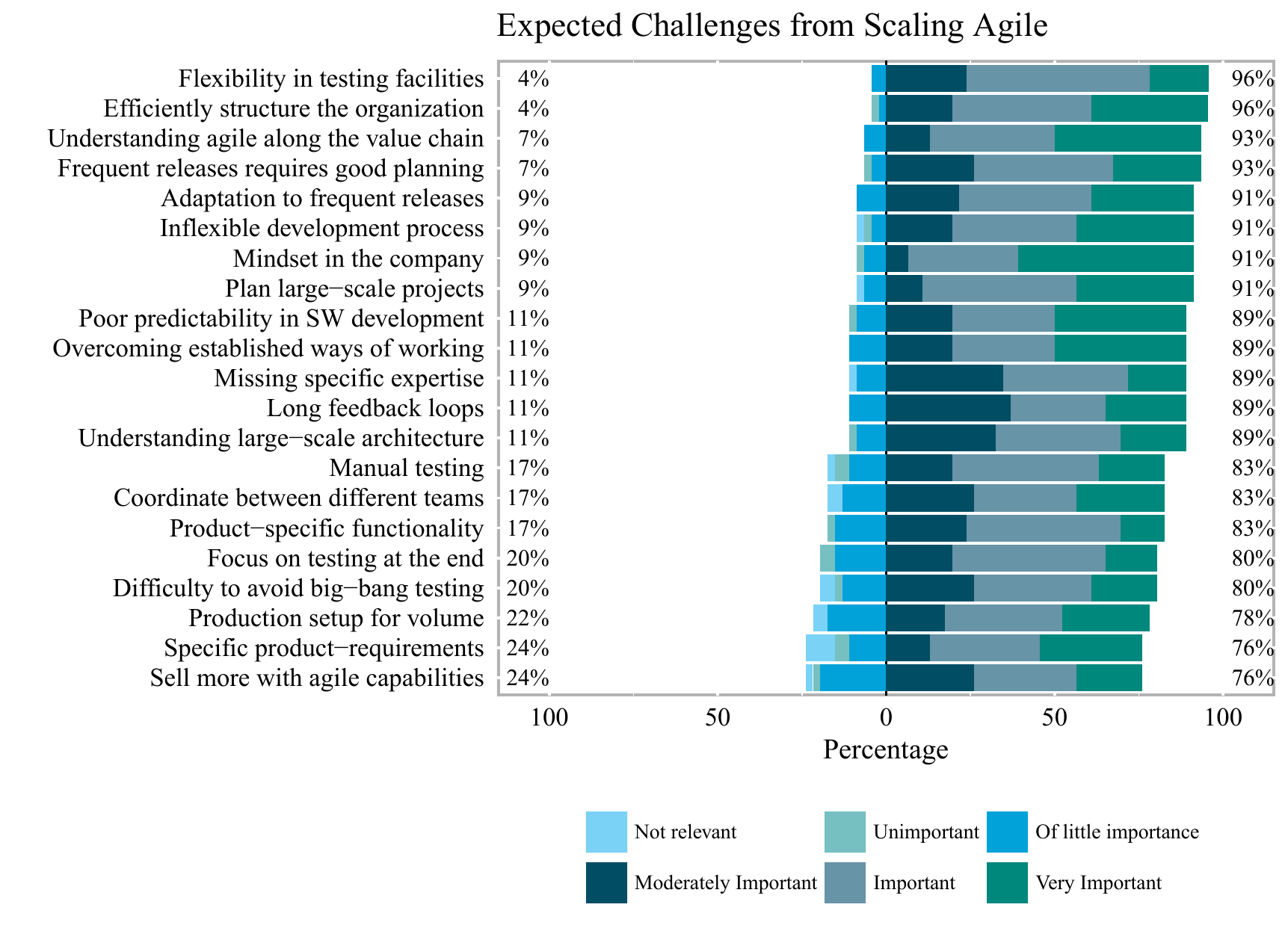}
\caption{Foreseeable challenges when scaling agile beyond software development teams; participants expressed their feedback using an ordered response scale.}
\label{fig:challenges}
\end{figure*}

The first part of the study investigated the expected benefits and foreseeable challenges when scaling agile beyond individual teams. 
As this is primarily a qualitative self-assessment from the view of the participants, we chose an ordered response scale for collecting and presenting the data.
Figures \ref{fig:benefits} and \ref{fig:challenges} summarize the main findings therefrom; a detailed analysis and discussion of the findings were reported in \cite{berger_expectations_2015}.
While the primary benefit of successful scaling of agile is expected higher quality, the main challenge was lack of flexibility in testing facilities. 
This is in line with the detailed analysis of the individual statements as shown in Tab.~\ref{tab:distributionBenefitsChallenges}.

Examining further our qualitative data, we identified a clear pattern that these expectations were mentioned because agile practices should enable quicker and better feedback to developers compared to previous practices. From our data, we could identify three feedback loops that run in parallel in these types of organizations.
The inner loop is similar to the single team sprint \emph{sprint} in Scrum. The middle loop differs from single team agile methods  since it is not planned \emph{releases}, as in e.g. XP, but  aims to coordinate the work of multiple teams.

\begin{enumerate}  
\item The local loop, which is performed within a single team, typically consisting of module or component development and testing every sprint.
\item The integration loop of software integrated with hardware and mechanics and the associated  verification \& validation.
\item The customer feedback loop.
\end{enumerate}

\begin{figure}[htb!]
\centering
\includegraphics[width=.8\linewidth]{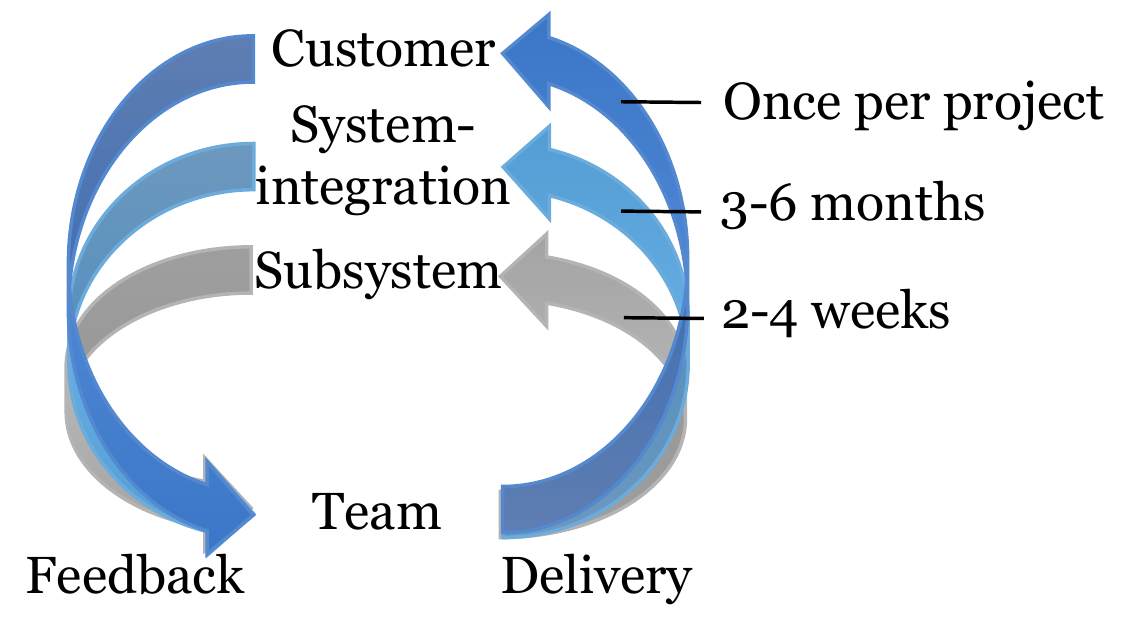}
\caption{Large-scale agile development has three feedback loops with different contexts and results, each having a different cycle time.}
\label{fig:loops}
\end{figure}

These three loops have different time scales as shown in Fig.~\ref{fig:loops}.
Just optimizing these feedback loops would contribute to a number of other agile benefits.
Typical key performance indicators (KPI) would be how many cycles of every loop is possible
to fit within a project; for the studied companies, five times would be a typical number for
the middle loop in their present projects.

\subsection{RQ-2: Key Practices and Goals in Mechatronic Organizations}
\label{sec:practices}
The first phase of the study unveiled that efficient integration is the key enabler in the context of scaling agile, and since this must be solved, the study focused on identifying goals and enabling agile practices that support accelerated
integration. As a result we identified 108 goals and practices in the second phase of the study. While the majority of them overlap with well-known best practices already established in agile and lean software development, there are, however, 26 goals and practices, as seen in Tab.~\ref{tab:practices}, that we 
identified as being unique to the mechatronics domain.

We organized the practices in Tab.~\ref{tab:practices} according to the maturity matrix from Stojanov et al.~\cite{stojanov_maturity_2015} with five maturity
levels of agile collaboration ranging from low to high: Collaborative, Evolutionary, Effective, Adaptive, and Encompassing. 
The levels serve as an evolutionary path through the defined stages supporting an organization attempting to scale agile development. Collaboration is considered an essential agile value and is therefore the \nth{1} level. The \nth{2} level is to develop software through an evolutionary approach. The  \nth{3} level is  to  effectivley and efficiently develop high quality software. The next level is using multiple levels of feedback to respond to change. The final \nth{5} level is to achieve an all-encompassing environment to sustained agility.
The matrix also sorts practices according to
agile principle, as derived from Beck et al.~\cite{beck_manifesto_2001}.
The items in the lower part of this matrix can be considered as topics that need to be addressed at a company towards unlocking the potential for continuous delivery/continuous deployment/continuous experimentation \cite{holmstrom_olsson_climbing_2012}.

Finally, we indicate per statement how many companies gave input to our individual interviews that support it (column ``A'') and how many companies from the control set (column ``B'') mentioned topics supportive to it; we sorted the statements based on their respective level of support from the participating companies in a descending level.

15 of the practices in Tab.~\ref{tab:practices} are related to \emph{technical excellence}, which is not surprising since the selection was on practices where the mechatronics domain differs from pure software in terms of complexity and constraints, i.e.~technical differences. 

The level of support indicated in columns ``A'' and ``B'' shows that 16 statements are supported at least by one company from the cases set and one from the control set.
Furthermore, 13 of these statements (i.e., more than 80\%) are supported by a majority of the case companies indicating a strong support for these agile principles particular to the mechatronics domain.

Below we will elaborate on the 16 practices confirmed by our control group.

\begin{table*}
\centering
\caption{
26 goals and practices particular to mechatronics development for scaling agile. The items are mapped to the Agile maturity level as suggested by Stojanov et al.; column ``A'' indicates how many different companies stated this in the individual interviews (range: 0\%, 25\%, 50\%, 75\%, and 100\%) and column ``B'' indicates during how many interviews with interviewees from the control set we found statements of the particular category (range: 0\%, 50\%, and 100\%). Statements that could not be confirmed from both groups were separated per maturity level. 
}
\label{tab:practices}
\begin{tabular}{p{9.0cm}p{1.9cm}p{4.9cm}|p{0.03cm}p{0.03cm}}
\hline
\textbf{Description} & \textbf{Agile maturity level} & \textbf{Agile principle} & A & B\\
\hline
Allow for SW deployment after production & 5 Encompassing  & Customer Collaboration & \pie{180} & \pie{0} \\
\hline
\hline
Minimize the number of point of contacts between SW, HW and mechanics & 4 Adaptive  & Technical Excellence &  \pie{360} & \pie{180} \\
Reduce variant complexity (component level) & 4 Adaptive  & Technical Excellence &  \pie{270} & \pie{180} \\
Allow for integrations of not the full product (e.g. Simulations) & 4 Adaptive  & Technical Excellence &  \pie{270} & \pie{180} \\
Not using the same planning/project gates for HW and SW & 4 Adaptive  & Plan and Deliver Software Frequently & \pie{270} & \pie{180} \\
Reduce variant complexity (product level) & 4 Adaptive  & Plan and Deliver Software Frequently &  \pie{90} & \pie{180} \\
\hline
Reduce variant complexity (product families level) & 4 Adaptive  & Embrace Change to Deliver Customer Value & \pie{0} & \pie{0}\\
Front-loading of the development process to stream-line industrialization is avoided & 4 Adaptive  & Plan and Deliver Software Frequently & \pie{90} & \pie{0} \\
\hline
\hline
Do not isolate disciplines & 3 Effective  & Human Centricity &  \pie{360} & \pie{360} \\
Do not depend on manual deployment & 3 Effective  & Technical Excellence &  \pie{360} & \pie{180} \\
Integration is a continuous activity (every 4 weeks) & 3 Effective  & Technical Excellence &  \pie{270} & \pie{360} \\
Move towards platforms & 3 Effective  & Technical Excellence &  \pie{270} & \pie{180} \\
Move complexity from mechanics to software / moves lead-time & 3 Effective  & Technical Excellence &  \pie{180} & \pie{180} \\
\hline
SW development is allowed to deliver a new release to the production every sprint & 3 Effective  & Technical Excellence &  \pie{360} & \pie{0} \\

Avoid the need to involve suppliers & 3 Effective  & Human Centricity &  \pie{180} & \pie{0} \\
Target SW is put as last on the HW in production & 3 Effective  & Technical Excellence &  \pie{0} & \pie{0}\\
\hline
\hline
Minimize supplier lead-times & 2 Evolutionary  & Human Centricity &  \pie{270} & \pie{180} \\
Speedy deployment of test software to the (prototype) product & 2 Evolutionary  & Technical Excellence &  \pie{270} & \pie{180} \\
\hline
Do not modify off-the-shelf products & 2 Evolutionary  & Plan and Deliver Software Frequently &  \pie{180} & \pie{0} \\
Identify the Minimum Viable Product to do SW integration & 2 Evolutionary  & Technical Excellence &  \pie{180} & \pie{0}\\
\hline
\hline
Quick and dirty HW available to test SW functionality & 1 Collaborative  & Technical Excellence &  \pie{360} & \pie{180} \\
SW available to use in tests of HW development & 1 Collaborative  & Technical Excellence &  \pie{270} & \pie{360} \\
Multidisciplinary teams & 1 Collaborative  & Human Centricity &  \pie{270} & \pie{360} \\
Having an agile process to adjust technical interfaces & 1 Collaborative  & Human Centricity &  \pie{180} & \pie{180} \\
\hline
Simplify technical interfaces & 1 Collaborative  & Technical Excellence &  \pie{180} & \pie{0} \\
Have cross-disciplinary and joint documentation & 1 Collaborative  & Technical Excellence &  \pie{0} & \pie{180} \\
\hline
\end{tabular}
\end{table*}

\subsubsection{Collaborative maturity level}
A set of \emph{technical interfaces} is usually defined by the system architecture, and these can only be changed through a process with all concerned stakeholders. This process must be as agile as the individual teams in order to achieve agility at scale.
Truly agile teams should consist of people from \emph{multiple different disciplines}; software, hardware, and mechanical engineering, and not only be cross-functional in the
sense they can handle different tasks such as coding, testing etc.
It is crucial to have \emph{quick and dirty hardware available to test software} functionality as soon as the software is available. Similarly there must be relevant \emph{software available when testing hardware}.

\subsubsection{Evolutionary maturity level}
Outsourcing parts of the system to suppliers is quite common, typical through traditional contractual negotiation, which may be counterproductive to being agile. If outsourcing is done it should be combined with actively working on \emph{minimizing suppliers lead-times}. 
A risk with self-governed teams is that developers can be isolated, especially if they are specialists or domain experts. Provide organizational structures that \emph{do not isolate developers}.
Deployment of software to the cluster of prototype products can be time consuming and care should be taken to have \emph{speedy deployment of test software to the prototypes}.

\subsubsection{Effective maturity level}
To speed up the integration in prototype products it is critical to facilitate it by e.g.~\emph{automation of software deployment} at scale instead of relying on complicated
manual procedures.

The simplest practice to understand, but maybe one of the most complex in practice, is to achieve \emph{full integration of software, hardware, and mechanics at least every four weeks}.
While more frequent integrations would favor software development, it seems to have only marginal benefit for the other disciplines. 
Achieving this pace of integration will drive many other changes towards agile practices.
We suggest a KPI on the transformation towards large-scale agile development based on the number of such integrations being done in a project of a certain length, for example going from six integrations in a 3-year project (i.e., two times per year) to 36 integrations in the same period (i.e. once per month).

Software development has shorter lead-times than hardware and mechanical development, therefore \emph{move complexity from hardware to software} if possible by careful system design.
Using common \emph{platforms to develop} multiple products in a product family speed up the development and facilitates adding and modifying features.

\subsubsection{Adaptive maturity level}
\emph{Reduce variant complexity} as a practice was seen in three different contexts. 
First, a large number of variation points tends to introduce technical complexity that in itself makes changes to both the code-base as well as the physical product anything but agile.
Second, in the context of planning we found examples when a new feature variant is planned to be introduced in a product project
subsequent projects tend to take for granted that the particular variant is \textit{freely} available without any significant development effort, thus constraining the possibility to reprioritize or remove it from the first project.
In practice the plan of introducing product variants in a product project that is planned to be ``carried over'' to other projects tend to reduce the flexibility in re-prioritization.
Third, 
deriving variants of existing products is a quick way to offer new products by embracing change. However, if this is not coupled with a strategy to reduce the number of variants to be maintained by R\&[]D compared to reducing variants that is offered to customers, the amount of work quickly grows. 

The practice of using the \emph{same planning/project gates} for hardware and software should be avoided due to the different nature of hardware and software development, the typical situation is to slow down software development to fit with the pace of hardware and mechanics.

Planning integrations based on the scope of the full product and the assumption that all teams can deliver against that scope limits agility. The setup must \emph{allow for integrations of an incomplete product}, e.g.~by replacing some software
or hardware parts with simulations.
The system architecture should strive to \emph{minimize number of point of contacts between software, hardware and mechanics} to simplify project coordination.

\subsubsection{Encompassing maturity level}
At this level we could only identify a precursor for continuous deployment: The unconfirmed practice of establishing an infrastructure that \emph{allow for software deployment also after the product has left the manufacturing plant}.


\section{Discussion and Related Work}
\label{sec:related}

In this section, we are discussing our findings in the context of related work and the validity or our results.


\subsection{Relation to Existing Evidence}

\subsubsection{Discussion about Benefits and Challenges}

Tab.~\ref{tab:dikert} lists the most important challenges found in our study. The immediate conclusion is that only the first challenge is specific to the mechatronic domain. A direct comparison with 35 challenges for large-scale agile transformations based on a systematic literature review (SLR) by Dikert et al.~\cite{dikert_challenges_2016} shows that challenges in agile transformations in different domains have more in common than what is different. The keywords used in their SLR explicitly focus on software by excluding papers containing \emph{manufacturing} in title or abstract. 

\begin{table}[t]
\centering
\caption{Comparison of challenges for large-scale agile in mechatronics domain and pure software.}
\label{tab:dikert}
\begin{tabular}{p{4.2cm} p{3.7cm}}
\hline
\textbf{Mechatronics challenges} & \textbf{Dikert et al.~\cite{dikert_challenges_2016}} \\
\hline
Flexibility in testing facilities & \emph{No equivalent} \\
Efficiently structure the organization & Internal silos kept \\
Understanding agile along the value chain  & Misunderstanding agile concepts \\
Frequent releases requires good planning  & Challenges in adjusting to incremental delivery pace \\
Adaptation to frequent releases  & Challenges in adjusting to incremental delivery pace \\
Inflexible development process  & Using old and new approaches side by side \\
Mindset in the company  & General resistance to change \\
Plan large-scale projects  & Challenges in adjusting product launch activities \\
Poor predictability in SW development  & \emph{No equivalent} \\
Overcoming established ways of working  & Skepticism towards the new way of working \\
Missing specific expertise  & Internal silos kept \\
Long feedback loops  & Challenges in adjusting to incremental delivery pace \\
Understanding large-scale architecture & Achieving technical consistency \\
\hline
\end{tabular}
\end{table}

It is interesting to note that the primary challenge identified in our study, \emph{Fexibility in testing facilities}, was not commonly seen in comparable recent papers. The only found example is Eliasson et al.~\cite{eliasson_agile_2014} describing Virtual test environments used in two cases at one of the companies presented in this study.

Scaling agile requires an \emph{efficient structure of the organization}, this is confirmed by e.g.~Dikert et al.~\cite{dikert_challenges_2016} who mention the challenge of when ``internal silos are kept'' keeping specialized knowledge inside internal boundaries. Könnölä et al.~\cite{konnola_agile_2016} mention the overall challenge of
``the specialization of the team members in the hardware or software tasks'', which confirms our identified practices concerning the importance of coordinated multidisciplinary work in agile development.

\emph{Understanding agile concepts along the value chain} is difficult in the mechatronics domain, where the value for the customer traditionally appears in the transaction of delivering the product. Wallin et al.~\cite{wallin_integrating_2002} discuss the mapping between software development life-cycle models, such as stage-gate or agile, and business decision models, and gives an example from a mechatronic company showing that this is not a new problem. However it does not seem to be solved since unfamiliarity with agile concepts in organization is still seen, as evident in the misunderstanding of agile mentioned by Dikert et al.~\cite{dikert_challenges_2016}. 

Another major challenge is the \emph{adjustment to frequent releases}, both from a
planning perspective, but also from a delivery perspective. The first main challenge in
using agile methods in embedded system development mentioned by 
Könnölä et al.~\cite{konnola_agile_2016} is the ``the slow nature of the hardware
development''. This confirms several of our results displaying the different paces of
software and hardware development and the need to address this when doing large-scale
agile development in the mechatronics domain. Examples of challenges mentioned by Könnölä et
al.~are \emph{inflexible development process}, \emph{long feedback loops},
\emph{understanding large-scale architecture}, and \emph{product-specific functionality}. Dikert et al.~\cite{dikert_challenges_2016} also mention the challenges to adjust to an incremental delivery pace. The challenge of adjusting the delivery pace is not unique to mechatronic systems, but may be even more accentuated in this domain.  However, there are examples of how future engineers already at university can learn to use agile concepts in development of mechatronic systems~\cite{klein_agile_2016}, with the support of agile meetings, artifacts, roles, and visualization methods.

It is interesting to see that most common challenge in SLR of Dikert et
al.~\cite{dikert_challenges_2016}, ``agile difficult to implement'', was not commonly
mentioned in our study, in spite of the mechatronics domain being potentially more
challenging compared to pure software systems. We cannot discern from our data if
this is because of an optimistic view of the agile transformation within the
organizations, or if there is a general hope that things will be smoother with
more frequent feedback.

The similarity between challenges found in this study and those found in literature
confirms our results, suggesting that the challenges are applicable also to other companies.
A smaller part of the identified challenges are unique to the embedded or mechatronics domain, but the challenges of scaling agile are mostly common to those of pure software systems.

\subsubsection{Discussion about key enabling Practices}
Our confirmed practices at the Collaborative level,
\emph{Having an agile process to adjust technical interfaces}, 
\emph{Quick and dirty HW available to test SW functionality}, and
\emph{SW available to use in tests of HW development}, are correspond to the more general practice ``Create team-driven agile practices inside the iterations'' prescribed by Könnölä et al.~\cite{konnola_agile_2016}. They recommend that ``Organize work planning according to disciplines, but make sure that the work of different disciplines is aligned and understood.'' which contrasts to our practice of \emph{Multidisciplinary teams} at this level. However they agree with our practice of \emph{Do not isolate disciplines} on the Effective level.

Our study identified two practices mitigating problems when outsourcing/subcontracting work with the first, \emph{minimize supplier lead-times}, at the Evolutionary level.
This is inline with the recommendations by Turk et
al.~\cite{turk_limitations_2002} stating that agile contracts for outsourcing needs to be flexible with respect to ``the requirements and deliverables that can vary within the boundaries defined'' explicating the need to tailor outsourcing in agile projects.
The other practice at the Evolutionary level, \emph{Speedy deployment of test software to the (prototype) product}, is also within the recommendation of ``Create team-driven agile practices inside the iterations'' by Könnölä et al.~\cite{konnola_agile_2016}.

We identified four confirmed practices regarding Technical Excellence at the Effective level: \emph{Do not depend on manual deployment} is a common practice in all agile development regardless of domain, as is \emph{Integration is a continuous activity}
(cf.~e.g.~\cite{stojanov_maturity_2015, stahl_modeling_2014}). 
\emph{Move towards platforms} is mentioned as an advanced agile technique by Könnölä et al.~\cite{konnola_agile_2016} for organizations to select where appropriate. \emph{Moving complexity from hardware to software} seems to be a new practice identified in this study. 

At the Adaptive level there are four practices that partly relate to the design and implementation of the architecture, which is not a well-researched area according to Yang et al.~\cite{yang_systematic_2016}, even less so for mechatronic systems.
The discrepancy in development pace between software and hardware
is already mentioned by \cite{konnola_agile_2016} as ``consensus between the quickly changing software and slower hardware development''. However we identified the importance of \emph{not} solving this by the simple approach to slow down software to the pace of hardware.

Many agile practices seem to be applicable across domains, even if we identified a number of practices that are unique to the mechatronics domain. However the selection of large-scale agile practices to be used in a certain organization still has to be tailored to the context. This only confirms the need identified by Dikert et al. to study large-scale agile frameworks in general~\cite{dikert_challenges_2016} and especially their applicability to the mechatronics domain.

\subsection{Threats to Validity}
\label{sec:ThreatsToValidity}

Here, we are discussing potential threats to the validity to our study according to
Runeson and Höst (cf.~\cite{runeson_guidelines_2009}). We described in 
Sec.~\ref{sec:ValidityProcedure} our procedure to aim for validity during data collection and processing.

\subsubsection{Construct Validity} 
The main threat to construct validity is if our selection of case companies are not representative of the meachatronics domain, and if their attempts of scaling agile development is not relevant to the research questions in Section~\ref{sec:rq}.
All six participating companies have a long development history of more than 30 years developing, manufacturing and selling mechatronic systems, many are also market leaders in their respective product domain. 
The companies participating in the Software Center all have the desire to become more agile to better meet their respective market's needs, and those participating in our study share the need of transforming their organization to become more agile to remain competitive. Thus, the case selection is both relevant and is able to answer the research questions.

A second threat to construct validity is if the  respondents to the survey perceive the difference between the negative categories of the ordered response scale to be smaller than the difference between the positive categories. However, the survey result is mainly used to rank the benefits and challenges and not  making any qualitative statements.

\subsubsection{Internal Validity}
Our data collection procedure 
face the risk of reflecting opinions from participants about aspects of internal processes that do not work today in general without being necessarily linked with transforming towards agile principles.
Therefore, for both, the on-site workshops and the surveys started with a brief overview of the principles stated in the Agile Manifesto to establish a common understanding. 
Here, we see a potential risk as the participants from non-software department had a chance for clarification during the workshops while we had not control if participants in the online survey had a similar understanding of agile principles.
The respective companies' points-of-contact also played an important role to compose participants for the on-site workshops where we had only limited influence to state what expected role would contribute best.
We aimed for mitigating this risk by explicitly letting survey participants forward the link to more colleagues who they feel could also contribute, i.e. snow-balling.
Furthermore, we tried to reduce this risk by involving several companies that face similar challenges in their organizational transformation to 
consolidate our data set.

\subsubsection{External Validity/Generalizability} The potential threat to validity in our study concerning generalizability originates from the number of participating companies. While we collaborated with six partners, we tried to reduce this potential risk by using two of them as controls
to confirm our findings.
We also deliberately included companies from different domains
to avoid reporting 
practices that might be only linked to a particular business domain. 
Furthermore, found related work suggest our findings are similar to existing evidence while we also complement the body-of-knowledge with results particular to the mechatronics domain.

\subsubsection{Reliability} Our overall study design aimed for reliability in the data collection and analysis steps, as described in Section~\ref{sec:method}. To avoid bias from an individual researchers both researchers have been involved in all data collection and analysis steps. Furthermore, we also involved the respective companies' points-of-contacts to allow them for giving feedback for our preliminary findings and to reduce the risk of being dependent on only the authors conducting this study.

\section{Conclusions and Future Work} 
\label{sec:conclusion}

108 goals and practices related to large-scale agile development were identified in the study, of these only a quarter were unique to the mechatronics domain. This leads to the conclusion that much of the research about in scaling agile can be considered valid regardless of the application domain. 
The large set of identified practices shows that there is still no silver bullet in accomplishing this, it is rather the tuning of a large number of interacting practices that determines the success of scaling agile in companies with already established ways-of-workings.
However, we identified 16 key practices special to mechatronics development 
considered key-enablers when scaling agile development beyond single software teams. Most of these are concerned with continuous or regular integration, and associated verification and validation between software, hardware and mechanics.

\subsection{Impact/Implications}

Improved quality was the number one perceived benefit of successfully scaling agile
in a mechatronics company, and to achieve this it is necessary to streamline the
integration between software, hardware, and mechanics. We can summarize how to achieve
this as two main objectives for industrial practitioners:

\begin{enumerate}
\item Transform your organization so the cycle time for full integrations between software, hardware, and mechanics is no longer than four weeks. A shorter cycle time (i.e.~continuous integration on a product level) benefits software development, but not necessarily other disciplines. A suggested KPI for improvement would be: How many full integrations are completed in a project of a certain length?

\item Ensure that the right feedback based on this is received promptly by the concerned stakeholders. A suggested KPI for improvement would be: How long (in days) does it take between ``a team made a delivery'' until ``the team receives the integration test results involving their part''?
\end{enumerate}

Most of the practices identified are enablers to these two objectives and their achievement will drive implementation of many other practices, accelerating the transformation towards scaled agile in the mechatronics domain.

\subsection{Limitations}

This study on companies that have an established way of conducting R\&D and a long history for manufacturing mechatronic products that become more and more software-intense.
We studied companies that are at the edge of rolling out agile principles from their software teams to the non-software neighbors to unveil key enabling practices to facilitate an adoption of agile. As the different companies have different transformation paces, we do not have further data at this point in time about their individual success journeys. Furthermore, we did not study young companies that might have a different culture in their current way of approaching engineering challenges (i.e. companies being agile ``from day one'').

\subsection{Future Work}

In future studies, some of the companies might be revisited again to continue monitoring their transformation process. Thereby, one might unveil further parameters that we did not see clearly at this point in time; furthermore, one might observe parameters that accelerate adopting agile in non-software environments while others might be perceived as concurrent in their nature.


\section*{Acknowledgment}
We are grateful to the companies who significantly supported this study in the context of Software Center.


\bibliographystyle{IEEEtran}
\bibliography{2017ICSE.bib}

\end{document}